\documentclass{mem}
\usepackage{natbib}
\usepackage{txfonts}
\usepackage{balance}
\usepackage{graphicx}
\usepackage[a4paper,breaklinks,pdftex]{hyperref}
\idline{75}{282}
\begin{document}
\def\teff{$T\rm_{eff }$}
\def\kms{$\mathrm {km s}^{-1}$}

\title{On the relative distance of Magellanic Clouds using Cepheid NlR and Optical-NIR PW relations}

   \subtitle{}

\author{
L. \,Inno\inst{1,2},  
G.~Bono\inst{1,3},
N.~Matsunaga\inst{4},
M.~Romaniello\inst{2},
F.~Primas\inst{2}, 
R.~Buonanno\inst{1,5},
F.~Caputo\inst{3},
K.~Genovali\inst{1},
C.D.~Laney\inst{6,7},
M.~Marconi\inst{8},
\and A.~Pietrinferni\inst{5}         
}

  \offprints{L. Inno}

\institute{Dipartimento di Fisica, Universit\`a di Roma Tor Vergata, 
via della Ricerca Scientifica 1, 00133 Rome, Italy; \email{laura.inno@roma2.infn.it}
\and European Southern Observatory, Karl-Schwarzschild-Str. 2, 85748 Garching bei Munchen, Germany
\and INAF--OAR, via Frascati 33, Monte Porzio Catone, Rome, Italy
\and Kiso Observatory, Institute of Astronomy, School of Science, The University of Tokyo,
10762-30, Mitake, Kiso-machi, Kiso-gun,3 Nagano 97-0101, Japan
\and INAF-Osservatorio Astronomico di Collurania, via M. Maggini, 64100 Teramo, Italy
\and Dept. of Physics and Astronomy, N283 ESC, Brigham Young University, Provo, UT 84601, USA
\and South African Astronomical Observatory, P.O. Box 9, Observatory 7935, South Africa
\and INAF-Osservatorio Astronomico di Capodimonte, via Moiarello 16, 80131 Napoli, Italy}

\authorrunning{Inno}

\titlerunning{On the relative distance of Magellanic Clouds}

\abstract{We present new estimates of the relative distance of the Magellanic Clouds (MCs) by using NIR and Optical-NIR Cepheid Period Wesenheit (PW) relations. The relative distances are independent of uncertainties affecting the zero--point of the PW relations, but do depend on the adopted pivot periods. We estimated the pivot periods for fundamental (FU) and first overtone (FO) Cepheids on the basis of their period distributions. We found that $\log P$=0.5 (FU) and $\log P$=0.3 (FO) are solid choices, since they trace a main peak and a shoulder in LMC and SMC period distributions. By using the above pivot periods and ten PW relations, we found MC relative distances of 0.53$\pm$0.06 (FU) and 0.53$\pm$0.07 (FO) mag. Moreover, we investigated the possibility to use mixed-mode (FU/FO, FO/SO) Cepheids as distance indicators and we found that they follow quite well the PW relations defined by single mode MC Cepheids, with deviations typically smaller than 0.3$\sigma$. 
}
\maketitle{}

\section{Introduction}

The cosmic distance ladder is a fundamental step not only for astrophysics, 
but also for observational cosmology \citep{mould11,suyu12}.
The quest for precise and accurate cosmic distances involved a 
paramount observational and theoretical effort to constrain the 
systematics affecting distances based either on geometrical methods 
or on distance indicators \citep{madore10,storm11a, matsunaga11,bono13} 
In this context a relevant breakthrough has been the very accurate 
and precise measurement of the distance to the Large Magellanic Cloud (LMC)
recently performed by \citet{pietrzynski13}. They used eight 
double eclipsing binary systems including two red giants distributed 
across the LMC bar and found a true distance modulus 
of 18.493$\pm$0.008 (statistical)$\pm$0.047 (systematic) mag. 
The precision of the distance is at the 2\% level and it is the most precise distance 
ever obtained of an extragalactic stellar system.  The new measurement 
will have a substantial impact not only on the new estimation of the 
Hubble constant, but also to validate other standard candles.   

The above issues are mainly dealing with absolute distance determinations. 
The absolute distances based on one of the variants of the Leavitt's law --i.e. 
Period-Luminosity (PL), Period-Luminosity-Color (PLC),  
Period-Wesenheit (PW) relations-- do require a detailed analysis of the dependence of both the 
zero-point and the slope on chemical composition and reddening
\citep{romaniello08, bono10, ngeow12, ripepi12, inno13}.  
The same applies to the many variants of the Baade-Wesselink 
method and their dependence on the parameter to transform radial 
velocities into pulsation velocities \citep{nardetto04,storm11a,groenewegen13}.
The literature concerning pros and cons of the different distance 
scales is significant \citep{alves04,feast08,bono08,walker12}.  

Relative distances have the indisputable advantage that they are independent 
of the absolute zero-point. These distances are quite useful not only for 
the cosmic distance scale, but also to constrain possible systematic 
uncertainties affecting different distance indicators.

\section{Photometric data and theoretical models}
\begin{figure}[t!]
\includegraphics[height=0.40\textheight, width=0.5\textwidth]{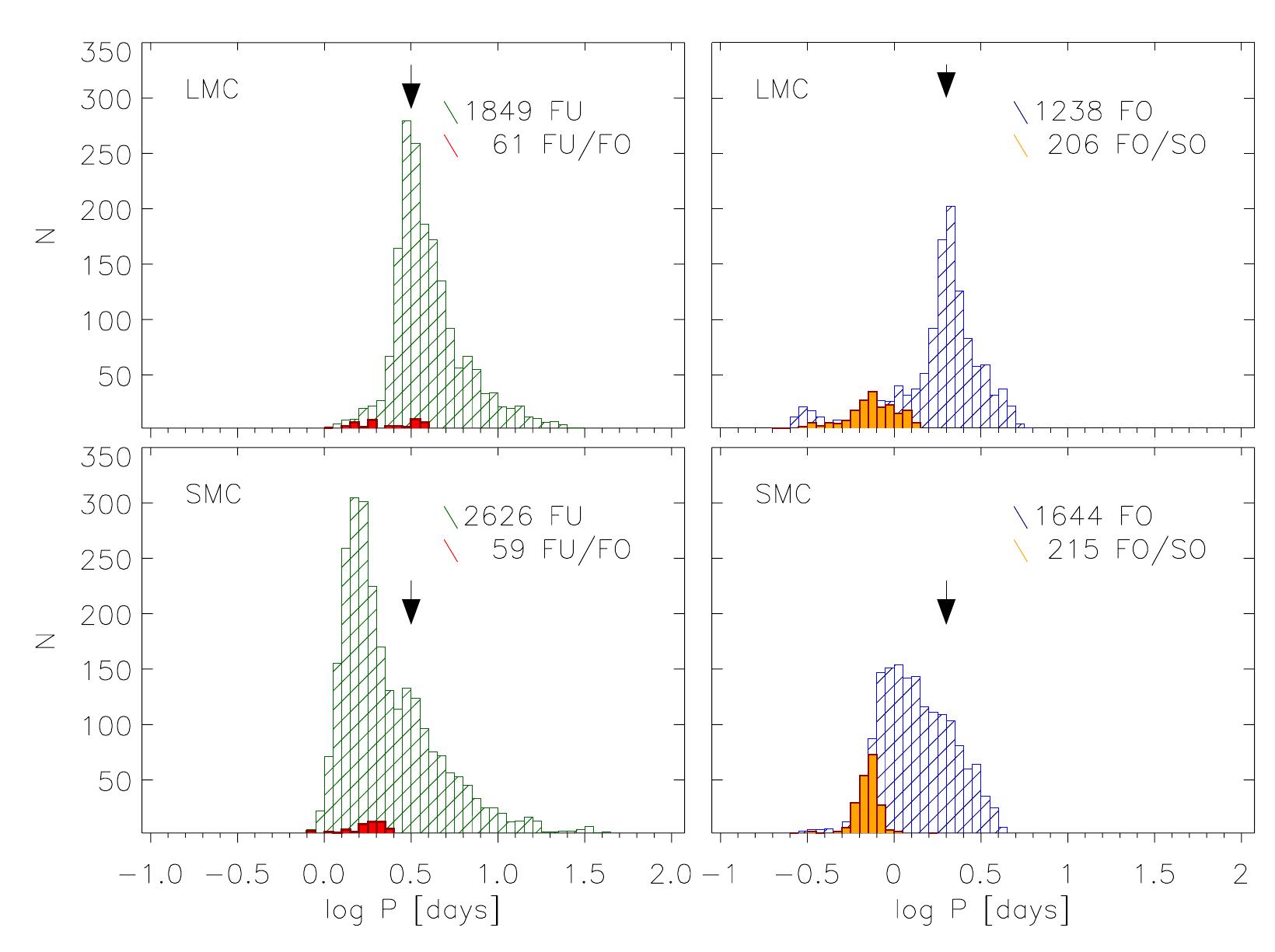}
\caption{\footnotesize
Left --Period distributions for FU (green), FU/FO (red) of the LMC (top) and SMC (bottom) Cepheids.
Right --The same for  FO (blue), FO/SO (orange) LMC (top) and SMC (bottom) Cepheids.}
\label{f0}
\end{figure}

\begin{figure}[t!]
\includegraphics[height=0.40\textheight, width=0.5\textwidth]{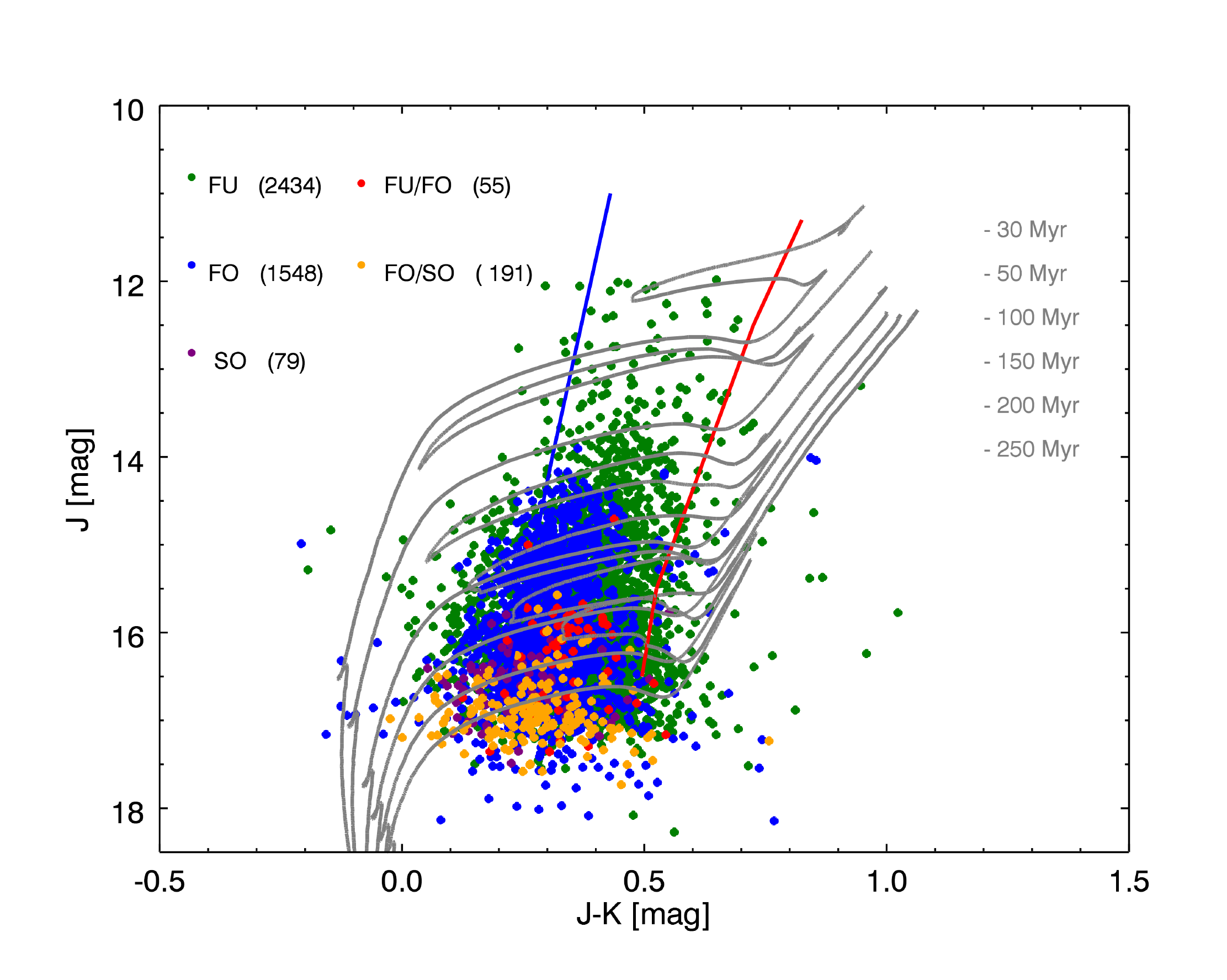}
\caption{\footnotesize
NIR --$J$,$J$-$K_{\rm S}$--Color-Magnitude Diagram (CMD) of 
 FU (green dots),  FO (blue dots), SO (purple dots) and double-mode
 (FU/FO, red dots; FO/SO, orange dots) SMC Cepheids. 
The grey lines display stellar isochrones with ages ranging 
from 30 Myr to 250 Myr (see labeled values), based on non-canonical 
evolutionary models (see Section 3). They are plotted by assuming 
a true distance modulus of $\mu$=18.93 mag,   
a mean reddening of $A_V$= 0.18 mag and the reddening law by \citet{cardelli89}
The cool (red) and the hot (blue) edge of the Cepheid IS are also shown \citep{marconi05}.   
}
\label{f1}
\end{figure}

The OGLE-III micro--lensing survey \citep[][2010]{sos2008} 
collected the most complete catalog of MC Cepheids in the Optical bands ($V,I$).
The OGLE-III Catalog of variable stars (CVS) includes $V$,$I$ band light curves for 
$\sim$3300 Cepheids in the LMC  and  $\sim$4500 Cepheids 
in the SMC.
The Fourier decomposition of these light curves allowed \citet{sos00} to properly identify 
the different pulsation modes -- LMC: 1848 fundamental-mode (FU), 1228 first-overtone (FO), 
61 fundamental-first-overtone double-mode (FU/F), 
203 first-second overtone double-mode (FO/SO) pulsators; 
SMC: 2626 FU, 1644 FO, 59 FU/FO and 215 FO/SO. 
The single epoch measurements for $\sim 90\%$ 
of the OGLE-III Cepheids were extracted from the IRSF/SIRIUS Near-Infrared (NIR: $J$,$H$,$K_{\rm S}$)
Magellanic Clouds Survey Catalog \citep{Kato07} and transformed into the 2MASS photometric system \citep{carpenter01}.
We adopted the $V$ and $I$ mean magnitudes, the $V$-band amplitude, the period $P$, and the
pulsation phase provided by the OGLE-III CVS to transform the FU Cepheids single-epoch NIR data to the NIR mean magnitudes by applying the template light curve, as described by \citet{templ}.
We also included the NIR mean magnitudes by \citet{persson04} for 41 long-period  FU Cepheids in the LMC.
We ended up with a sample of $\sim$3300 LMC and $\sim$4400 SMC Cepheids 
with $VIJHK_{\rm S}$ bands photometry, including double-mode pulsators 
(LMC: 1840 FU, 1202 FO, 60 FU/FO, 199 FO/SO; SMC: 2587 FU, 1579 FO, 55 FU/FO,195 FO/SO).
In our analysis we focused on FU and FO Cepheids, in order to evaluate the PW relations \citep{inno13}.
To enlarge the statistics of our sample, we now include the double-mode pulsators.
For these Cepheids, we decided to adopt the period of the dominant mode.
The period distributions of FU, FU/FO and FO, FO/SO Cepheids for the LMC (top) and SMC (bottom) are shown in Fig.~\ref{f0}. The period distributions of FU (green dashed bars) and FU/FO  (red filled bars) LMC Cepheids in the left--top panel cover the same period range and show similar shape. The same applies for the FU and FU/FO SMC Cepheids in the left-bottom panel of Fig.~\ref{f0}.
Moreover, the period distributions of FO (blue dashed bars) and FO/SO (orange filled bars) MC Cepheids in the right--top (LMC) and right--bottom (SMC) panel of Fig.~\ref{f0} also show similar shape. 
The period distribution of Cepheids plays a crucial role to constrain the recent 
Star Formation History (SFH) of their host galaxies \citep{becker77,alcock99,antonello02}.
Cepheids are indeed good tracers of the young stellar populations, 
since their evolutionary status is well established. 
They are intermediate-mass stars evolving along the blue loop 
in the helium-core and hydrogen-shell burning stage. 
They obey to well known Period-Age and Period-Color-Age relations,
 which indicate that younger Cepheids have longer periods.
Thus, the period distribution traces the underlying stellar population and
several factors can affect its shape: the initial mass function, the metallicity, 
the star formation history, the mass loss rate and the topology of the instability strip.
To further constrain the evolutionary status of MC Cepheids, 
Fig.~\ref{f1} shows the NIR ($J,J-K_{\rm S}$) 
color--magnitude diagram (CMD) of the  SMC Cepheids. 
The FO Cepheids (green circles), with longer periods, 
are brighter and bluer compared with FU Cepheids
(blue circles).  
The gray lines show a set of isochrones with a scaled-solar chemical 
mixture and a chemical composition ($Z$=0.004,$Y$=0.256) typical of SMC young population \citep{pietrinferni04,pietrinferni06}\footnote{\tt http://albione.oa-teramo.inaf.it/}.
The isochrones are based on evolutionary models that account 
for mild convective core overshooting during the central hydrogen-burning phase 
and for a canonical mass-loss rate ($\eta=0.4$). 
The comparison between theory and observations indicates that 
SMC Cepheids have ages ranging from a few tens to a couple of hundred Myr. 
The blue loops for the older ($t\approx250$ Myr) and the younger ($t\approx 30$ Myr) 
isochrones do not extend over the whole color range of the Cepheids observed.  
This well known theoretical problem is related to the physical mechanisms 
and the input physics adopted in the evolutionary calculations \citep{bono00,neilson11,prada12}.
The vertical blue and red lines show the predicted edges of the
instability strip (IS) for the modal stability of both FU (red edge) and FO (blue edge)
Cepheids with similar metallicity \citep{marconi05}. 
We also included the double-mode pulsators.
The FU/FO Cepheids (red dots) are found in the same range of color of the 
FU Cepheids, but at lower luminosities, as expected. The FO/SO Cepheids (orange dots)
are also fainter and older than the FO ones, as expected.
Fig.~\ref{f1} shows a quite good agreement between theory and observations.
Moreover, we have to take into account that the theory predictions are obtained for a fixed mean
chemical composition and are not corrected for differential reddening or geometrical effects.

%


\section{MC Relative Distance}
\begin{figure}[t!]
\includegraphics[height=0.40\textheight, width=0.5\textwidth]{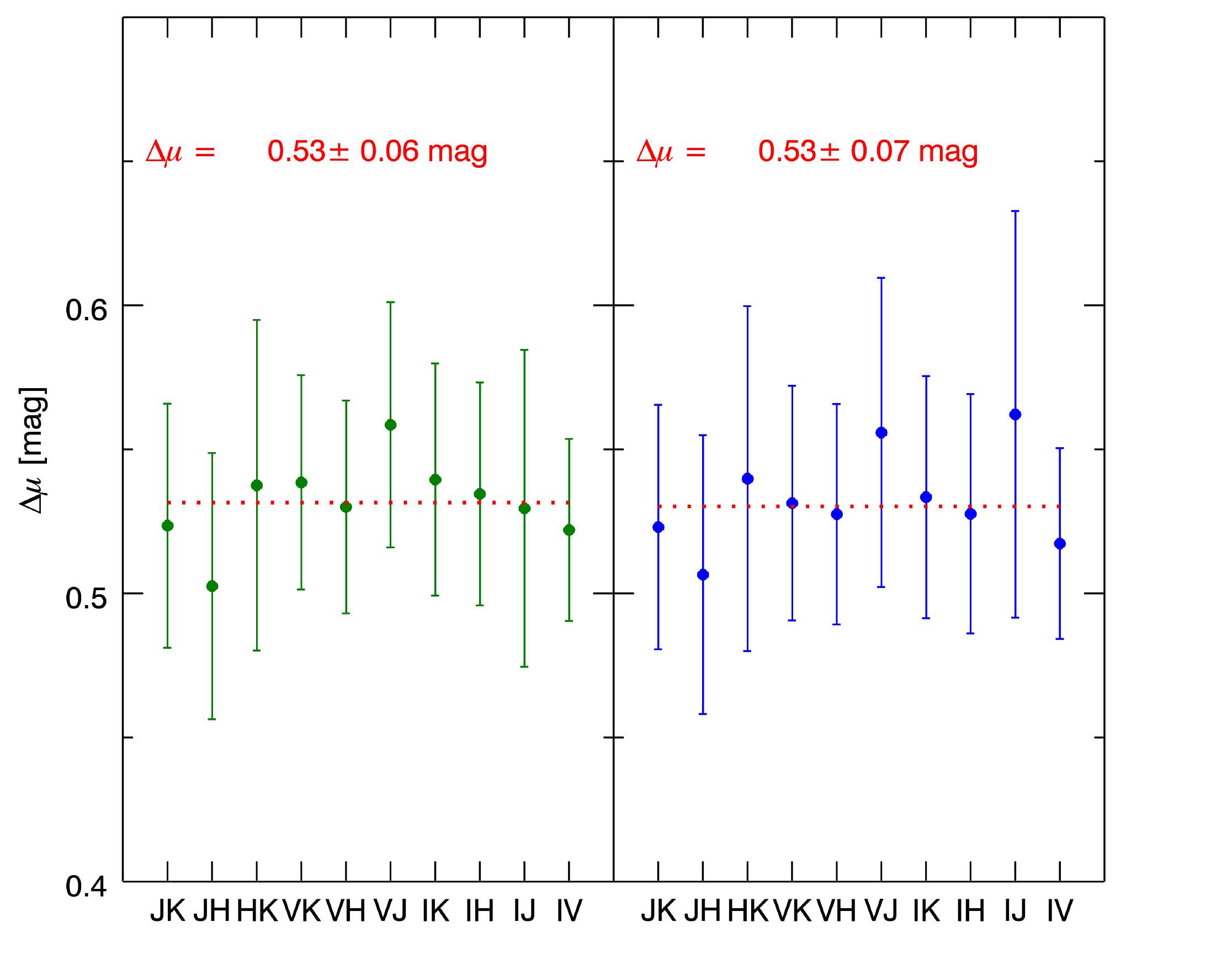}
\caption{\footnotesize
Left --Comparison between the MC relative distances for the ten different PW relations 
for FU Cepheids, evaluated at the pivot period $\log P$= 0.50.
The error bar for each point accounts for both the dispersion 
and the error on the coefficients of the PW relations.
The red dashed lines is the weighted average of the ten values, 
that is also labeled in the top. 
Right --Same as the left, but for FO Cepheids. 
The relative distances are evaluated at the pivot period $\log P$= 0.30 (see Section~3)}
\label{f2}
\end{figure}

\begin{figure}[t!]
\includegraphics[height=0.40\textheight, width=0.5\textwidth]{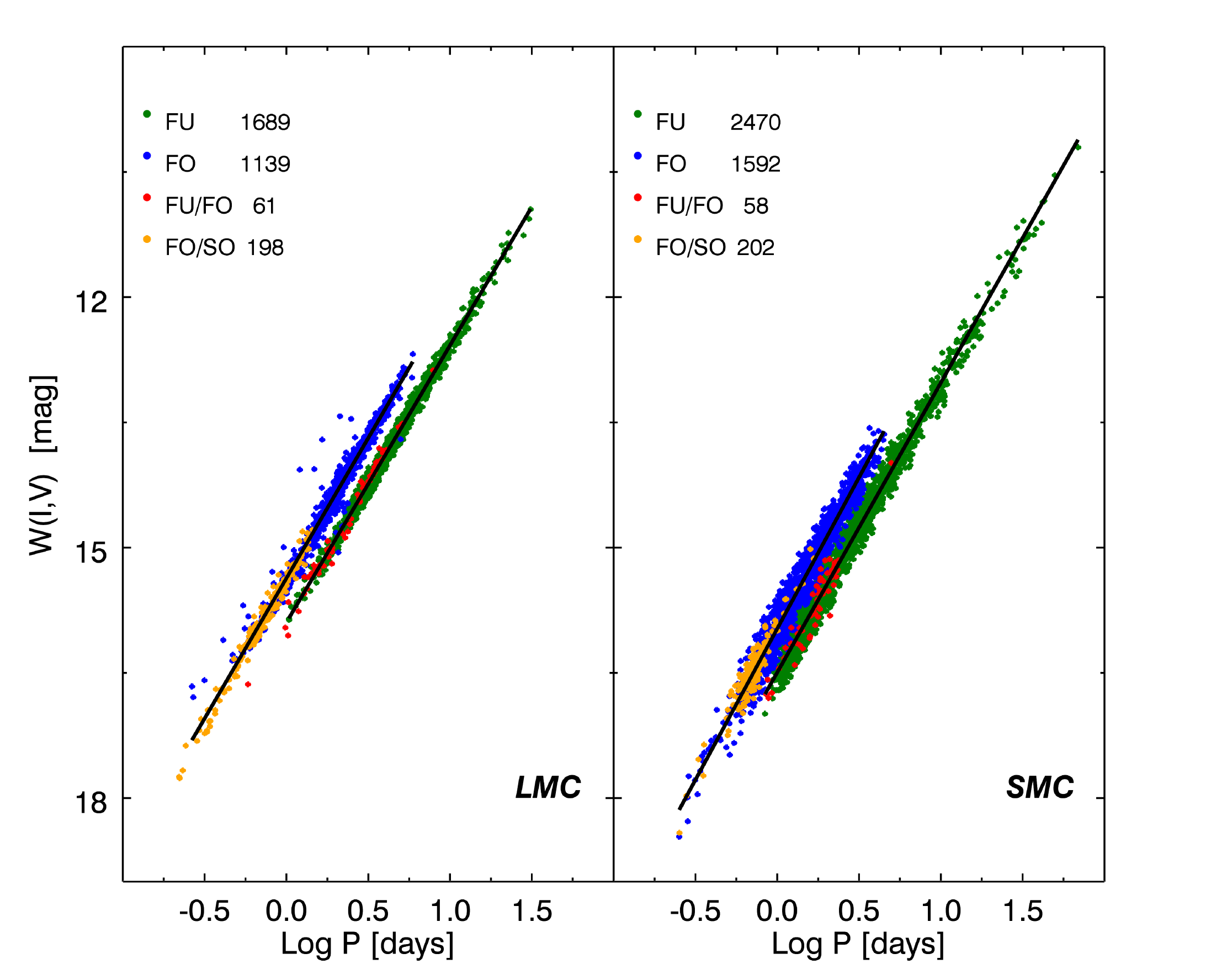}
\caption{\footnotesize
Left -- $(V,I) $ PW relations for LMC Cepheids: FU (green dots),  
FU/FO (red dots), FO (blue dots) and FO/SO (orange dots) Cepheids.
The solid lines show the linear fits for FU and FO Cepheids.
Right --Same as the left, but for the SMC.}
\label{f3}
\end{figure}

In  order to address the problem concerning the linearity of the PW relations, 
in \citet{inno13} we devised a new empirical test based on the relative
distance between SMC and LMC. 
In Fig.~\ref{f2} we compare the MC relative distance 
modulus estimated with the ten different PW relations
for FU (left) and FO (right) Cepheids.
We computed these distances according to the following equation:
$\Delta\mu= a_{SMC}-a_{LMC}+\log P \ (b_{SMC}-b_{LMC})$;
where $a,b$ are the coefficients of the FU and FO PW for the SMC and LMC  
and $\log P$ is the fixed pivot period.
The choice of the pivot period is somewhat arbitrary in literature,
ranging from $\log P$=0.5\citep{groenewegen00} or $\log P$=0.8 \citep{devau78}
to $\log P$=1.0 \citep{udalski99,freedman01} or $\log P$=1.3 \citep{matsunaga11}.
However, with a glance to the period distribution in the left panels of Fig.~\ref{f0} 
it is clear that the PW relations mainly depend on the bulk of the Cepheid distribution and 
the number of Cepheids that constrains the PW relation at $\log P$=1.0 is quite poor 
compared with the number of Cepheids at the peak of the distribution (LMC:$\log P$=0.5; SMC; $\log P$=0.3). 
The peak of the period distribution of the FU LMC Cepheids, indicated by the black arrows in the left panels of Fig.~\ref{f0}, 
is a more suitable pivot period,since the SMC period distribution also shows 
a substantial number of Cepheids for the same period range.
The choice of a new pivot period ($\log P$= 0.5) allow us 
to correctly sample the PW relation.
The same applies for the LMC FO peak distribution at $\log P$= 0.3, 
indicated by the black arrows in the right panels of Fig.~\ref{f0}.
By using the ten PW relations for FU and FO Cepheids, given in \citet{inno13}, we evaluated ten different MC relative distance moduli, shown in Fig.~\ref{f2}. The error bars are drawn taking into account both
the error on the coefficients $a,b$ of the PW relations and the statistical dispersion associated to the linear fit.
In the case  of the FO Cepheids, the error bars are larger, because of the larger dispersion in the FO PW relations.
This is due to the lack of a template for these pulsators, as discussed in \citet{inno13}.
We also computed the weighted average of these ten values, to reduce the associated 
error, and we found $\Delta\mu= 0.53 \pm 0.06$ mag for the FU PW relations and  $\Delta\mu= 0.53 \pm 0.07$ mag for the FO PW relations.
The two results are in very good agreement, confirming the robustness of the relative distance estimation.
Fig.~\ref{f3} shows the comparison between the distribution of the double-mode pulsators in 
the PW($V,I$) diagram and the optical PW relations (solid lines) presented in \citet{inno13}
for the single-mode FU (green dots) and FU (blue dots) Cepheids in the LMC (left) and SMC (right).
Interestingly enough, the FU/FO (red dots) Cepheids follow the FU PW relation, 
with a deviation that is on average smaller than 0.3$\sigma$ for the LMC and
0.1$\sigma$ for the SMC Cepheids, where $\sigma$ is the statistical
dispersion associated to the fit.
The FO/SO (orange dots) Cepheids follow the FO PW relation with a deviation that is
similar or smaller (0.2$\sigma$, LMC; 0.3$\sigma$, SMC). 
We also tested the difference using the NIR and Optical-NIR PW relations and we found that the deviation is even smaller ($<$0.05 $\sigma$).
The above findings indicate that the Optical and NIR PW relations can be adopted to estimate individual distances of the mixed-mode pulsator.

\section{Conclusions} 
We present new estimates of the relative distance of the MCs by using both 
NIR and Optical-NIR Cepheid PW relations. The main advantage in using relative 
distances is that they are independent of uncertainties affecting the zero--point 
of the PW relations. The pivot periods to estimate the relative distance was 
estimated on the basis of the period distribution. We found that $\log P$=0.5   
and $\log P$=0.3 are solid choices for FU and FO Cepheids, since they mark the primary
peaks in the LMC period distribution and a shoulder in the SMC period distribution.   
By using the above pivot periods and the ten Cepheid PW relations, we found relative distances 
of 0.53$\pm$0.06 (FU) and 0.53$\pm$0.07 (FO) mag. 

Current findings indicate that both FU and FO Cepheids can  be safely 
adopted to estimate relative distances. The above results further support 
the hypothesis that the difference in the absolute distance of MCs 
based on FU and FO Cepheids \citep{inno13} is mainly caused by 
the limited accuracy of the absolute zero--point of FO PW relations.   
Indeed, the zero-point relies on a single object --Polaris \citep{vanleeuwen07}--since 
we still lack accurate trigonometric parallaxes for other nearby 
Galactic FO Cepheids. 

We also investigated the possibility to use mixed-mode (FU/FO, FO/SO) Cepheids 
as distance indicators using both NIR and optical-NIR PW relations. We found that 
they follow quite well the PW relations defined by single mode MC Cepheids. 
Indeed, the deviations are typically smaller than 0.3$\sigma$. 
\begin{acknowledgements}
We are indebted to the editor, P. Bonifacio, for his constant support.
Two of us (L.I. and K.G.) thank the SAIT for partial support to attend the EWASS conference.
One of us (G.B.) thanks ESO for support as a science visitor.
This work was partially supported by PRIN-INAF 2011 ``Tracing the
formation and evolution of the Galactic halo with VST" (P.I.: M. Marconi)
and by PRIN-MIUR (2010LY5N2T) ``Chemical and dynamical evolution of
the Milky Way and Local Group galaxies" (P.I.: F. Matteucci).
\end{acknowledgements}

\bibliographystyle{aa}

\begin{thebibliography}{}
\bibitem[Alcock et al.(1999)]{alcock99} Alcock, C., Allsman, R.~A., Alves, D.~R., et al.\ 1999, \aj, 117, 920
\bibitem[Alves(2004)]{alves04} Alves, D.~R.\ 2004, New Astronomy Reviews, 48, 659 
\bibitem[Antonello et al.(2002)]{antonello02} Antonello, E., Fugazza, D., \& Mantegazza, L.\ 2002, \aap, 388, 477 
\bibitem[Becker et al.(1977)]{becker77} Becker, S.~A., Iben, I., Jr., \& Tuggle, R.~S.\ 1977, \apj, 218, 633 
\bibitem[Bono et al.(2010)]{bono10}  Bono, G., Caputo, F., Marconi, M., \& Musella, I.\ 2010, \apj, 715, 277 
\bibitem[Bono et al.(2008)]{bono08} Bono, G., Stetson, P.~B., Sanna, N., et al.\ 2008, \apjl, 686, L87 
\bibitem[Bono et al.(2000)]{bono00} Bono, G., Castellani, V., \& Marconi, M.\ 2000, \apj, 529, 293 
\bibitem[Bono et al.(2013)]{bono13} Bono,G., Matsunaga, N., Inno, L.,Lagioia, E.P., \& Genovali, K. \ 2013, in Cosmic-Ray in star-forming environments, ed. D.F. Torres,ÊO. Reimer (Sant Cugat Forum in Astrophysics; Berlin: Springer), Êin press
\bibitem[Cardelli et al.(1989)]{cardelli89} Cardelli, J.~A., Clayton, G.~C., \& Mathis, J.~S.\ 1989, \apj, 345, 245 
\bibitem[Carpenter(2001)]{carpenter01} Carpenter, J.~M.\ 2001, \aj, 121, 2851 
\bibitem[de Vaucouleurs(1978)]{devau78} de Vaucouleurs, G.\ 1978, \apj, 223, 730 
\bibitem[Feast et al.(2008)]{feast08} Feast, M.~W., Laney, C.~D., Kinman, T.~D., van Leeuwen, F., 
\& Whitelock, P.~A.\ 2008, \mnras, 386, 2115 
\bibitem[Freedman \& Madore (2010)]{madore10} Freedman, W.~L., \& Madore, B.~F.\ 2010a, \araa, 48, 673 
\bibitem[Freedman et al.(2001)]{freedman01} Freedman, W.~L., Madore, B.~F., Gibson, B.~K., et al.\ 2001, \apj, 553, 47 
\bibitem[Groenewegen (2013)]{groenewegen13} Groenewegen, M.~A.~T.\ 2013, \aap, 550, A70 
\bibitem[Groenewegen (2000)]{groenewegen00} Groenewegen, M.~A.~T.\ 2000, \aap, 363, 901 
\bibitem[Inno et al.(2013)]{inno13} {Inno, L., Matsunaga, N., \&Bono, G.}, et al. 2012, \apj, 764, 84 
\bibitem[Kato et al.(2007)]{Kato07} Kato, D., Nagashima, C., Nagayama, T., et al.\ 2007, \pasj, 59, 615 
\bibitem[Marconi et al.(2005)]{marconi05} Marconi, M., Musella, I., \& Fiorentino, G.\ 2005, \apj, 632, 590 
\bibitem[Matsunaga et al.(2009)]{matsunaga09} Matsunaga, N., Feast, M.~W., \& Menzies, J.~W.\ 2009, \mnras, 397, 933 
\bibitem[Matsunaga et al.(2011)]{matsunaga11} Matsunaga, N., Feast, M.~W., \& Soszy{\'n}ski, I.\ 2011, \mnras, 413, 223 
\bibitem[Mould (2011)]{mould11} Mould, J.\ 2011, \pasp, 123, 1030 
\bibitem[Nardetto et al.(2004)]{nardetto04} Nardetto, N., Fokin, A., Mourard, D., et al.\ 2004, \aap, 428, 131 
\bibitem[Neilson et al.(2011)]{neilson11} Neilson, H.~R., Cantiello, M., \& Langer, N.\ 2011, \aap, 529, L9 
\bibitem[Ngeow (2012)]{ngeow12}Ngeow, C.-C. 2012, ApJ, 747, 50 (N12) 
\bibitem[Persson et al.(2004)]{persson04}Persson, S.~E., Madore, B.~F., Krzemi{\'n}ski, W., et al.\ 2004, \aj, 128, 2239 
\bibitem[Pietrinferni et al.(2004)]{pietrinferni04} {Pietrinferni, A., Cassisi, S., Salaris, M., et al.} 2004, \apj, 612, 168
\bibitem[Pietrinferni et al.(2006)]{pietrinferni06} {Pietrinferni, A., Cassisi, S., Salaris, M., et al.} 2006, \apj, 642, 797
 \bibitem[Pietrzy{\'n}ski et al.(2013)]{pietrzynski13} Pietrzy{\'n}ski, G., Graczyk, D., Gieren, W., et al.\ 2013, \nat, 495, 76 
\bibitem[Prada Moroni et al.(2012)]{prada12} {Prada Moroni, P.G., Gennaro, M., Bono, G., et al.} 2012, \apj, 749, 108
\bibitem[Ripepi et al.(2012)]{ripepi12} Ripepi, V., Moretti, M.~I., Marconi, M., et al.\ 2012, \mnras, 424, 1807
\bibitem[Romaniello et al.(2008)]{romaniello08} Romaniello, M., Primas, F., Mottini, M., et al.\ 2008, \aap, 488, 731 
\bibitem[Soszy{\'n}ski et al.(2005)]{templ} Soszy{\'n}ski, I., Gieren, W., \& Pietrzy{\'n}ski, G.\ 2005, \pasp, 117, 823 
\bibitem[Soszynski et al.(2000)]{sos00} Soszynski, I., Udalski, A., Szymanski, M., et al.\ 2000, AcA, 50, 451 
\bibitem[Soszy{\'n}ski et al.(2008)]{sos2008} Soszy{\'n}nski, I., Poleski, R., Udalski, A., et al.\ 2008, AcA, 58, 163 
\bibitem[Storm et al.(2011)]{storm11a} Storm, J., Gieren, W., Fouqu{\'e}, P., et al.\ 2011, \aap, 534, A94 
\bibitem[Suyu et al.(2012)]{suyu12} Suyu, S.~H., Treu, T., Blandford, R.~D., et al.\ 2012, arXiv:1202.4459 
\bibitem[Udalski et al.(1999)]{udalski99} Udalski, A., Szymanski, M., Kubiak, M., et al.\ 1999, AcA, 49, 201 
\bibitem[van Leeuwen et al.(2007)]{vanleeuwen07} van Leeuwen, F., Feast, M.~W., Whitelock, P.~A., \& Laney, C.~D.\ 2007, \mnras, 379, 723 
\bibitem[Walker(2012)]{walker12} Walker, A.~R.\ 2012, \apss, 341, 43 
\end{thebibliography}

\end{document}